\documentclass[a4paper]{article}

\usepackage{INTERSPEECH2022}
\usepackage{amsmath,graphicx}
\usepackage[hang,flushmargin]{footmisc}
\usepackage[textfont=it,tableposition=top]{caption}
\usepackage{hyperref}
\hypersetup{colorlinks=true}
\usepackage{multirow,enumitem}
\usepackage[hang,flushmargin]{footmisc}
\usepackage{lipsum,soul}

\newcommand\blfootnote[1]{%
  \begingroup
  \renewcommand\thefootnote{}\footnote{#1}%
  \addtocounter{footnote}{-1}%
  \endgroup
}

\renewcommand{\paragraph}[1]{\textit{#1}:\quad}

\title{A Universally-Deployable ASR Frontend for Joint Acoustic Echo Cancellation, Speech Enhancement, and Voice Separation}
\name{Tom O'Malley, Arun Narayanan, Quan Wang}
\address{Google LLC, U.S.A}
\email{\{omalleyt, arunnt, quanw@google\}.com}
\begin{document}
\ninept
\maketitle
\begin{abstract}
Recent work has shown that it is possible to train a single model to perform joint acoustic echo cancellation (AEC), speech enhancement, and voice separation, thereby serving as a unified frontend for robust automatic speech recognition (ASR). The joint model uses contextual information, such as a reference of the playback audio, noise context, and speaker embedding.  In this work, we propose a number of novel improvements to such a model. First, we improve the architecture of the \emph{Cross-Attention Conformer} that is used to ingest noise context into the model. Second,  we generalize the model to be able to handle varying lengths of noise context. Third, we propose \emph{Signal Dropout}, a novel strategy that models missing contextual information. In the absence of one or more signals, the proposed model performs nearly as well as task-specific models trained without these signals; and when such signals are present, our system compares well against systems that require all context signals. Over the baseline, the final model retains a relative word error rate reduction of 25.0\% on background speech when speaker embedding is absent, 
and 61.2\% on AEC when device playback is absent.

\end{abstract}

\vspace{0.05in}

\blfootnote{We thank A. Gruenstein, A. Park, J. Walker, N. Howard, and S. Panchapagesan for several useful discussions.}
\vspace{-0.1in}
\section{Introduction}
\label{sec:intro}

Neural network based end-to-end models~\cite{PrabhavalkarRaoSainathLiEtAl17,BattenbergChenChildCoatesEtAl17,HoriWatanabeZhangChan2017,li2021betterfaster}, large-scale training data~\cite{mirsamadi2017multi,hakkani2016multi,NarayananMisraSimPundakEtAl18}, and improved data augmentation strategies~\cite{kim2017mtr, park2019specaugment, medennikov2018investigation} have significantly improved the robustness of automatic speech recognition (ASR) systems. However, factors such as device echo from smart speakers, harsher background noise, and competing speech still significantly deteriorate performance~\cite{barker2017thirdchime,barker2018fifthchime}.
While it is possible to train separate ASR models that address each adverse condition, in practice, it is hard to maintain multiple task-specific ASR models and dynamically pick the model that perfectly suits a use case. Furthermore, with large scale multi-domain \cite{NarayananMisraSimPundakEtAl18} and multi-lingual modeling \cite{adams2019massively, pratap2020massively} gaining more interest, it is increasingly infeasible to optimize the ASR model to address varying use cases and adverse noise conditions simultaneously. Therefore, it is often convenient to use separate frontend feature-processing modules that address these adverse conditions, and to train and maintain them separately from the ASR models.

Adverse noise conditions can broadly be classified into 3 distinct categories: background noise, competing speech, and device echo.

\paragraph{Background noise} Non-speech background noise is handled via data augmentation strategies such as multi-style training (MTR) \cite{lippmann1987multi,kim2017mtr}. This technique is relatively straight-forward to apply during ASR training, and as a result, large-scale ASR models are generally robust to moderate levels of non-speech noise. However, background noise can still degrade performance in low signal-to-noise ratio (SNR) conditions~\cite{ko2017study}.

\paragraph{Competing speech} Competing speech refers to the presence of one or more non-target speakers in the utterance. ASR models are typically trained on a single speaker, since in multi-speaker conditions it is generally unclear as to which speaker the model should attend to, without additional contextual information. Models that separate out multiple speakers \cite{tripathi2020end} are generally sub-optimal since it is difficult to determine how many speakers to support. As a consequence, competing speech conditions are quite challenging for typical ASR models.

\paragraph{Device echo} With interactive devices such as smart home speakers, ASR performance degrades severely when the device is playing back audio, especially if that audio contains discernible speech, which is typical for voice assistants. Acoustic echo cancellation (AEC) techniques~\cite{hansler2005acoustic,benesty2001advances,zhang2018deep} address this issue. Signal processing \cite{hansler2005acoustic, benesty2001advances, benesty2011perspective, enzner2014acoustic} and neural network \cite{zhang2018deep, fazel2019deep, lei2019deep, ding2020textual} based solutions have both been proposed for AEC. What makes the task distinct from the others is that the reference signal of the device playback is usually available and can be used for noise suppression.

The aforementioned noise conditions have been addressed in the literature, typically, using task-specific models. A recent work has shown that it is possible to address all 3 types of interference simultaneously with a single model \cite{omalleyt2021joint}. 
Such a model uses multiple contextual signals:
\begin{itemize}[leftmargin=*]
    \item A noise context signal, which is a noise-only segment prior to the utterance, to help remove background noise.
    \item A speaker embedding vector for improving performance in conditions with competing speech.
    \item The playback reference signal for removing device echo.
\end{itemize}
The model proposed in \cite{omalleyt2021joint} makes the simplifying assumption that the necessary context signals are always present during training and inference. But in practice, not all contextual signals are available at all times. For example, speaker embedding vectors are usually obtained via an optional speaker enrollment process, and many device users choose to opt-out of this enrollment. Similarly, the reference signal for echo cancellation may not be available on certain devices, or are badly aligned and, therefore, unusable. In this work, we generalize this previous work to handle missing contextual information. Modeling missing information is a well studied area, especially in the audio-visual literature \cite{afouras2019my,zhang2019robust,yu2021audio}. These works typically deal with occlusion and corrupted signals, which is more common to visual signals. Unlike these earlier works, we specifically address completely missing contextual information and strictly for an acoustic frontend.

A related problem is the availability and the length of the noise context before the utterance. This depends on the existence and size of the audio buffer on the device, which can also vary widely. Apart from modeling missing information, we therefore also propose modifications to noise context modeling to allow for varying lengths of noise context. 

Finally, we propose improvements to the architecture of the Cross-Attention Conformer \cite{narayanan2021crossattention} that was used in \cite{omalleyt2021joint} for noise context modeling. 

The rest of this paper is laid out as follows: Sec.~\ref{sec:system} provides a detailed description of the proposed system. Sec.~\ref{sec:experiments} describes our general experimental settings. Experiments and results are presented in Sec.~\ref{sec:results} and conclusions in Sec.~\ref{sec:conclusion}.

\vspace{-0.1in}
\section{System}
\label{sec:system}

We start by giving a brief overview of the system in \cite{omalleyt2021joint}, which is what we build on, and then focus on the improvements that allow this system to generalize better.

\vspace{-0.1in}

\subsection{Overview}

\begin{figure}[tbh!]
	\centering
	\includegraphics[width=0.5\textwidth]{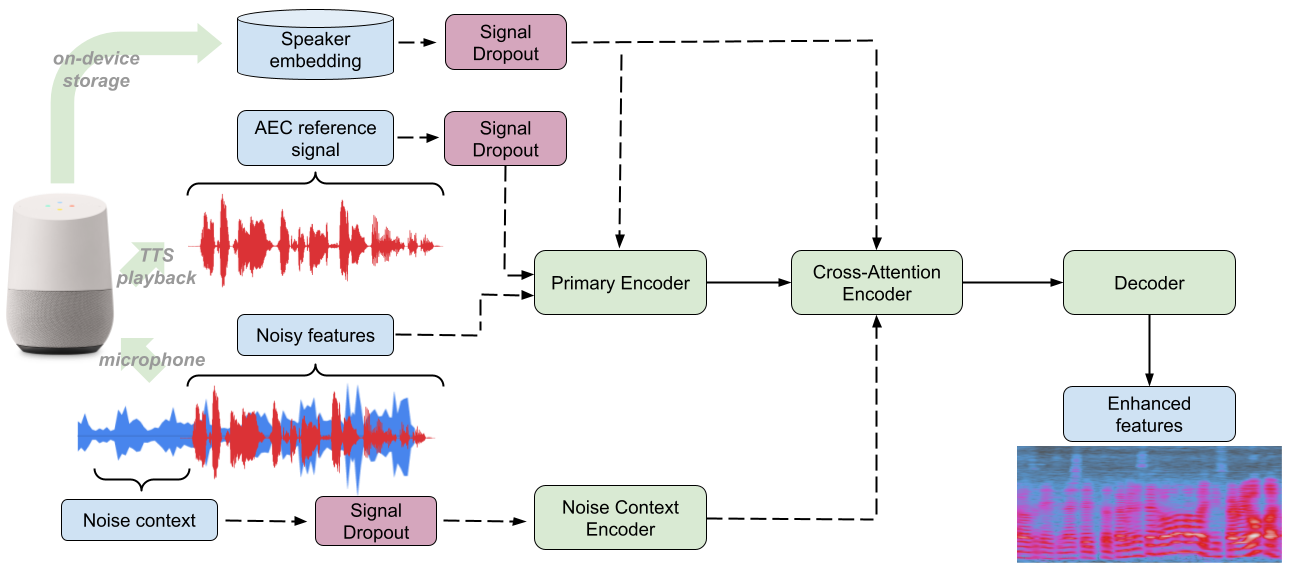}
	\caption{
	System overview. The system receives noisy features, and 3 optional signals: the device playback, noise context, and speaker embedding. Signal Dropout is performed only during training.
	}
	\label{fig:system_overview}
\end{figure}

Fig.~\ref{fig:system_overview} shows a block diagram of the overall system. The model is composed of a primary encoder, a noise context encoder, and a cross-attention encoder that combines the outputs of the two encoders. Each of these encoders consists of modified conformer blocks \cite{gulati2020conformer} that use a speaker embedding vector to modulate their input. Conformers have been shown to be well-suited for speech and audio tasks in prior work \cite{gulati2020conformer, chen2021continuous}. Modulation via a speaker embedding vector allows the model to focus on the speaker of interest during enhancement.

\vspace{-0.1in}

\noindent\paragraph{\bf{Primary Encoder}} The primary encoder consists of $N$ modified conformer blocks. The inputs to the primary encoder are log Mel-filterbank energy (LFBE) features from the noisy signal and the AEC reference signal, stacked together in the frame dimension. The encoder also receives a speaker embedding of the target speaker as a side input. At the start of each conformer block, the speaker embedding is combined with the block's inputs using feature-wise linear modulation (FiLM) \cite{perez2018film}. The modulated inputs are then passed to a standard conformer encoder.

\noindent\paragraph{\bf{Noise Context Encoder}}The noise context encoder is composed of $N$ standard conformer blocks, and is executed in parallel with the primary encoder. The encoder uses LFBE features from the noise context as input.

\noindent\paragraph{\bf{Cross-Attention Encoder}}The outputs of the primary encoder and the noise-context encoder are fed to the cross-attention encoder \cite{narayanan2021crossattention} as the input feature and the auxiliary feature, respectively. The cross-attention encoder is composed of $M$ modified cross-attention conformer blocks. Before each block, we repeat the process of combining the speaker embedding with the input features using FiLM, similar to the primary encoder. The cross-attention encoder summarizes encoded noise-context features with a cross-attention layer that uses the modulated input features as \emph{queries}. The summarized noise context is subsequently merged with the inputs using FiLM and a second cross-attention layer. In prior work, we show that the cross attention conformer is able to summarize highly non-stationary noise types, summarizing it separately for each input frame to be enhanced, thereby overcoming the limitations of using a single noise embedding to represent noise context \cite{narayanan2021crossattention}.

\vspace{-0.05in}

\subsection{Proposed improvements}
\subsubsection{Improvements to Cross-Attention Conformer}
\label{sec:cross-attention-conformer}

We modify the cross attention blocks to better summarize and merge the noise context with input features. After processing encoded input features and noise context features through feedforward and convolutional layers, we summarize the processed noise context features using cross-attention. Similar to \cite{omalleyt2021joint}, for cross-attention, the processed input features are used as queries, and the noise context features are used for deriving keys and values. But unlike \cite{omalleyt2021joint}, we skip the residual connection after cross-attention, relying on the subsequent FiLM layer to combine the noise-context summary with the input features. Following FiLM, we use a self-attention layer instead of the cross-attention layer used in \cite{omalleyt2021joint}.

Mathematically, if $x$, $m$, and $n$ are the encoded input, d-vector and the encoded noise context from the previous layer, the modified cross-attention encoder does the following:

\begin{align}
\begin{split}
\hat{x} &= x + r(m) \odot x + h(m),\\
\tilde{x} &= \hat{x} + \frac{1}{2}\textrm{FFN}(\hat{x}), \tilde{n} = n + \frac{1}{2}\textrm{FFN}(n),  \\
x^{\prime} &= \tilde{x} + \textrm{Conv}(\tilde{x}), n^{\prime} = \tilde{n} + \textrm{Conv}(\tilde{n}), \\
\color{blue} x^{\prime\prime} &\color{blue}= \textrm{MHCA}(x^{\prime}, n^{\prime}), \\
\color{blue} x^{\prime\prime\prime} &\color{blue}= x^{\prime} + x^{\prime} \odot r(x^{\prime\prime}) + h(x^{\prime\prime}), \\
\color{blue} x^{\prime\prime\prime\prime} &\color{blue}= x^{\prime\prime\prime} + \textrm{MHSA}(x^{\prime\prime\prime}), \\
y &= \textrm{LayerNorm}(x^{\prime\prime\prime\prime} + \frac{1}{2} \textrm{FFN}(x^{\prime\prime\prime\prime})). \\
\end{split}
\end{align}

Here, $\textrm{MHCA}$ and $\textrm{MHSA}$ stand for multi-headed cross attention and multi-headed self-attention, respectively. The modifications, compared to \cite{omalleyt2021joint}, are marked in blue. With these modifications, $x^{\prime\prime}$ now strictly corresponds to a noise summary, $x^{\prime\prime\prime}$ merges input features and noise summary, and $x^{\prime\prime\prime\prime}$ adds self-attention processing block on the combined features. $y$ is the final output of the processing block, which is passed on to the next layer, along with the d-vector, $m$, and the encoded noise context, $n$.

\subsubsection{Generalized Noise-context Modeling}
\label{sec:noise-context-length}

In the previous work, it was assumed that 6 seconds of background noise context was always available to the frontend model. In this work, we relax that assumption by training our model on randomly trimmed noise contexts, with the context length uniformly sampled from 0 to 6 seconds. The random length of the resulting noise context poses a challenge to the absolute positional embedding, which is commonly used in transformer / conformer blocks when modeling sequences. As the distance of each noise context frame from the start of the target utterance is now also dependent on the length of the noise context, adding the positional embedding \cite{vaswani2017attention,gulati2020conformer} during the pre-processing stage of the noise context can provide misleading information to both the noise context encoder and the cross attention encoder. For example, the last noise context frame will have different absolute positional embedding depending on the length of the noise context. Intuitively, it is the actual spectral information carried in the noise context (especially for non-stationary noise types) that helps improve enhancement. Furthermore, it is the distance of a noise context frame from the start of the target utterance that is more important than their absolute position within the noise context. 

Based on these intuitions, we consider two alternatives to absolute positional embedding. First, we drop the positional embedding in the noise context encoder, but use a reversed absolute positional embedding in the cross-attention encoder. That is, instead of using a positional embedding, $p(t)$, for the noise context frame, $n(t)$, the reversed positional embedding uses $p(T_N - t)$ as the embedding, where $T_N$ is the number of frames in the noise context. Therefore, the positional embedding encodes how long before the target utterance a frame of noise context occurred, fixing the misalignment problem otherwise introduced by the random trim. The second approach drops positional embedding altogether in both the noise context encoder and the cross-attention encoder. 

\subsubsection{Modeling Missing Context Signals}
\label{sec:missing-context-signals}

To allow the system to generalize to environments in which one or more context signals are missing, we propose using \emph{signal dropout} during training, wherein context signals are dropped out with a given probability. The idea is similar to dropout regularization \cite{srivastava2014dropout}, but applied at an input level. Our goal with signal dropout is to encourage the model to not just rely on the most relevant context signal, but to also utilize alternative context signals that may provide useful information for enhancing the input. For example, for AEC, even when the reference signal is missing, the speaker embedding vector will provide some information to separate the target signal from the noisy input.

Since the model architecture is typically static, the dropped-out context signal still needs to be presented in some fashion to the model. 
Therefore, when a context signal is dropped, we replace the features that would have been generated for that signal with all-zeros. For the reference signal for AEC, we thus create an all-zero feature of the same length and feature dimension as the utterance. The feature dimension is the same as that of the LFBE features used for the reference signal, if it were not missing. For the noise context signal, we create an all-zero feature with a length of 6 seconds, and the same dimension as the LFBE features. The speaker embedding is replaced by a 256-dimensional all-zero vector, which matches the length of the speaker embedding vector. All features are then fed, as before, to the joint ASR frontend. 

We also experimented with replacing the missing signals with a frame-level learned representation, rather than all-zeroes. However, we found that this did not improve performance over using all-zeros.
\begin{table}
  \centering
  \caption{Ablation study of the proposed Cross-Attention Conformer changes. WER are reported.}
  \begin{tabular}{|l|c|}

  \hline
  Model & Noise \\ \hline
  \hline
  
  Proposed Model & \bf 13.6 \\ \hline
  ~- Residual changes & 13.7 \\ \hline
  ~~~- Self-Attention changes \cite{narayanan2021crossattention} & 14.2 \\ \hline
  \end{tabular}
  \label{table:cross-attention-conformer}
\end{table}

\begin{table}[th]
  \centering
  \caption{Results using generalized noise context modeling. Reversed refers to reversed positional embedding for noise and None refers to no positional embedding. WER are reported.}
  \begin{tabular}{|c|cc|cc|}

  \hline
  Positional Embeddings & \multicolumn{2}{c|}{Speech} & \multicolumn{2}{c|}{Noise} \\ \hline
  & -5 dB & 5 dB & -5 dB & 5 dB \\
  \hline
  Reversed & 41.0 & 21.0 & 31.8 & 12.9 \\ \hline
  None & \bf 39.4 & \bf 20.4 & \bf 30.2  & \bf 12.3 \\ \hline

  \end{tabular}
  \label{table:random-trim}
  \vspace{-0.05in}
\end{table}

\begin{table*}[th]
  \centering
  \caption{Missing Context Signal Results. Word error rates are reported for -5 db SNR background speech, -5 dB SNR background noise, and -10 db SNR AEC. }
  \begin{tabular}{|c|ccc|c|c|c|}

    \hline
  Model &
  \multicolumn{3}{c|}{All context} &
  No dvector &
  No noise context &
  No playback \\ \hline

  & Speech & Noise & AEC &
  Speech & Noise & AEC \\ \hline

  Baseline & 69.2 & 36.5 & 80.5 & 69.2 & 36.5 & 80.5 \\ \hline

0\% Signal Dropout & \bf 44.6 & \bf 31.9 & \bf 23.9 & 61.7 & 34.9 & 65.5 \\ \hline

20\% Signal Dropout & 45.1 & 32.0 & 24.2 & 51.9 & \bf 33.6 & 31.2 \\ \hline

50\% Signal Dropout & 45.0 & 32.2 & 26.1 & \bf 50.9 & 34.3 & 30.6 \\ \hline

  Dedicated Model & - & - & - & 51.3 & 34.5 & \bf 29.0 \\ \hline

  \end{tabular}
  \label{table:missing-context}
  \vspace{-0.1in}
\end{table*}

\vspace{-0.1in}
\section{Experimental settings}
\label{sec:experiments}

\subsection{Datasets}

Similar to \cite{omalleyt2021joint}, we train on datasets derived from LibriSpeech~\cite{librispeech} as well as internal vendor-collected utterances. LibriSpeech consists of $281$k utterances. The vendor-collected sets consist of $1,916$k utterances. We treat these data sources as `clean' and add device echo, background noise, or competing speech to each utterance. The test sets are created using the test-clean subset of Librispeech.

\noindent\paragraph{Background noise} The speech enhancement training set is created using a room simulator \cite{kim2017mtr}. This first adds reverberation with T60s between 0 msec and 900 msec, then noise with an SNR in the range \mbox{[-10~dB, 30~dB]}. The noise snippets represent typical noise conditions such as kitchen, cars, etc. as well as publicly available noises from Getty\footnote{\url{https://www.gettyimages.com/about-music}} and YouTube Audio Library\footnote{\url{https://youtube.com/audiolibrary}}. Test sets are also created in a similar fashion, but using mixing conditions that are disjoint from training.

\noindent\paragraph{Competing Speech} In order to simulate multi-speaker conditions, we mix the training utterances with competing speech from the training datasets, chosen randomly. Test sets are also constructed in a similar fashion.

\noindent\paragraph{Device Echo} We use two types of device echoes \cite{howard2021neural}. The first type consists of entirely synthetic echoes, created using a reference signal played through a reverberant room simulator, with the speaker configured to be close to the microphone. Reference signals are drawn from Librispeech. The second type of training data contains re-recorded echoes using an internal TTS dataset. These utterances are re-recorded after being played back in a room on Google Home devices at varying signal levels. The goal is to capture microphone non-linearities, which are harder to model using a room simulator. The echoes are added to reverberant target speech signal at SNRs in the range \mbox{[$-20~$dB, $5$~dB]}. The AEC evaluation sets are created using only these re-recorded echoes, since that subset is both more challenging and closer to real-world conditions.

\vspace{-0.05in}
\subsection{Training details}

All models are trained in TensorFlow \cite{abadi2016tensorflow}, using the Lingvo \cite{shen2019lingvo} toolkit. All joint frontend models contain approximately 15M parameters. We use 2 conformer layers for each of the primary encoder, noise context encoder, and cross-attention encoder. Each conformer layer has 256 units, with the kernel size for convolution set to 15. 128-dimensional LFBE features are used for all signals, computed for 32 msec windows with 10 msec hop. The dvector is 256-dimensional. All attention layers use causal self-attention with a window of 65 frames in the past. 

The frontend model estimates an ideal ratio mask \cite{Narayanan2013IRM}, which is used to enhance the noisy LFBE features. The enhanced features are then passed to a pretrained recurrent neural net transducer based ASR model \cite{sainath2020streaming}. %
As losses, the frontend models use a combination of $\mathcal{L}_1$ and $\mathcal{L}_2$ distance between estimated and ideal masks, and an ASR loss that encourages the activations of a pretrained ASR encoder generated using clean and enhanced LFBE features to be close to each other.

\vspace{-0.05in}
\section{Results}
\label{sec:results}

All experiments are evaluated by passing noisy data through the enhancement frontend, and then passing the enhanced features to a pre-trained ASR model. We report the ASR model's word error rate (WER).

\vspace{-0.05in}
\subsection{Modified Cross-Attention Conformer}

We perform a few ablations on the improvements to the Cross-Attention Conformer. The results are shown in Table \ref{table:cross-attention-conformer}. We only report results on the evaluation data with background noise, since the Cross-Attention Conformer is primarily used to integrate noise context into the model. WERs are reported on an evaluation set consisting of background noise added to the dev-clean subset of Librispeech at a uniform range of SNRs from -5 dB to 15 dB. 
As can be seen, our proposed system achieves a reduction in relative WER of 4.2\% compared to the original Cross-Attention Conformer. Most of these gains are coming from replacing the last cross-attention with self-attention; moving the residual connection from the first cross-attention to the subsequent FiLM module improves WER by only $\sim$1\%.

\vspace{-0.05in}
\subsection{Generalized Noise-Context Modeling}

Table \ref{table:random-trim} shows the WERs using the various strategies to encode positional information for noise context. For these experiments, we train a model with no signal dropout on the training data with background noise and background speech.  Although using reversed positional embedding to encode the distance in time from the beginning of the target utterance is helpful, we find that removing the positional embedding entirely outperforms such a method, across all conditions. This is likely because it encourages the model to summarize noise context entirely based on the spectral information, as opposed to the position of the frame in the noise context.

\vspace{-0.05in}
\subsection{Modeling Missing Context Signals}

We train models with 2 different probabilities of dropping each context signal: 20\%, and 50\%. The same probability is used for each context signal. We also train dedicated models without speaker embedding, a model without reference signal, and a model without noise context, to use as strong comparison points for evaluating performance when a signal is missing. Similarly, we train a model with all contextual signals always present, to use as a strong comparison point for evaluating performance when all signals are present. The ``Baseline'' corresponds to using the noisy features directly for ASR without any enhancement.

For evaluations, we separate each noise type into its own evaluation set, and compare the performance of the model on each set at different SNRs. All results shown are for -5 dB SNR background speech, -5 dB SNR background noise, and -10 dB SNR acoustic echo settings. Additionally, to measure performance when a signal is absent, we create evaluation sets without speaker embedding, without noise context, and without the reference signal.

Table \ref{table:missing-context} shows the WER of the missing context signal experiments.
These results show that, when all context signals are present, the 20\% and 50\% signal dropout method performs nearly as well as a model trained without any signal dropout (0\% signal dropout). Additionally, when a context signal is missing, the models trained with signal dropout significantly outperform such a model and achieve comparable performance to a dedicated model trained entirely without that context signal. 

Training with a 20\% signal dropout rate appears sufficient to achieve the majority of these benefits. When all signals are present, the 20\% signal dropout model achieves a relative WER within 1.1\% on the background speech dataset, 0.3\% on the background noise dataset, and 1.2\% on the AEC dataset compared to the 0\% signal dropout model. Compared to dedicated models, the 20\% signal dropout model performs within 1.2\% relative WER on the background speech dataset when speaker embedding is missing, slightly better on the background noise dataset when noise context is missing, and within 7.6\% relative WER on the AEC dataset when device playback is missing.

Compared to baseline, when all signals are present the 20\% signal dropout model reduces relative WER in low SNR environments by 35.5\%, 12.3\%, and 70.0\%, on background speech, background noise, and AEC datasets, respectively. The model retains a relative reduction over baseline of 25.0\% on background speech when speaker embedding is absent, 7.9\% on background noise when noise context is absent, and 61.2\% on AEC when device playback is absent. We note that the model performs surprisingly well on AEC even when the device playback is missing, presumably because it is able to make use of the speaker embedding; this shows that the models trained with signal dropout learns to make use of the available information even when the most relevant context signal is missing.%

\vspace{-0.1in}
\section{Conclusion}
In this work, we presented an improved ASR frontend for joint AEC, speech enhancement, and voice separation. Our system can handle the presence or absence of multiple context signals, allowing it to be deployed in a wide variety of environments. In addition, we adapted the system to better incorporate noise context of varying lengths. Finally, we showed general-purpose architectural improvements to the originally proposed Cross-Attention Conformer. In future work, we plan to explore extending our approach to multi-channel inputs.

\label{sec:conclusion}

\newpage
\bibliographystyle{IEEEbib}
\bibliography{refs}

\end{document}